\documentclass[12pt,a4paper]{article}

\usepackage{latexsym}
\usepackage{amsfonts}
\usepackage{amssymb}
\usepackage{amsmath}
\usepackage[mathscr]{eucal}
\usepackage{theorem}
\usepackage{enumerate}
\usepackage{cite}
\usepackage{url}
\usepackage[dvips]{color}
\usepackage{graphicx}

\theoremstyle{plain}
\theorembodyfont{\itshape}
\newtheorem{thm}{Theorem}

\theorembodyfont{\upshape}

\newcommand{\mydate}{
 \ifcase\month \or
 January\or February\or March\or April\or May\or June\or
 July\or August\or September\or October\or November\or December\fi
 \space \number\year}

\newcommand{\smotimes}{\,{\scriptstyle \otimes}\,}

\newcommand{\smwedge}{{\scriptstyle \wedge\,}}

\newcommand{\upoast}{\ensuremath{%
 ^{\,\raisebox{0.1ex}{$\scriptscriptstyle \bigcirc$} \hspace{-0.45em} \ast}}}
\newcommand{\upostar}{\ensuremath{%
 ^{\raisebox{0.05ex}{$\scriptscriptstyle \bigcirc$} \hspace{-0.45em} \star}}}

\newcommand{\bwedge}{\raisebox{0.2ex}{${\textstyle \bigwedge}$}}

\begin{document}

\title{Hamiltonian Vector Fields on Multiphase Spaces \\
       of Classical Field Theory
       \thanks{Work partially supported by CNPq (Conselho Nacional de
               Desenvolvimento Cient\'{\i}fico e Tecno\-l\'o\-gico), Brazil}}
\author{Michael Forger$^{\,1}$~~and~~M\'ario Ot\'avio Salles$^{\,2}$
        \thanks{Work done in partial fulfillment of the requirements
                for the degree of Doctor in Science}}
\date{\normalsize
      $^1\,$ Departamento de Matem\'atica Aplicada, \\
      Instituto de Matem\'atica e Estat\'{\i}stica, \\
      Universidade de S\~ao Paulo, \\
      Caixa Postal 66281, \\
      BR--05315-970~ S\~ao Paulo, S.P., Brazil  \\[4mm]
      $^2\,$ Faculdade de Tecnologia de Carapicuiba \\
      Av.\ Francisco Pignatari, 650, \\
      BR--06390-310 Carapicuiba, S.P., Brazil
      }
\maketitle

\thispagestyle{empty}

\begin{abstract}
\noindent
 We present a classification of hamiltonian vector fields on multisymplectic
 and polysymplectic fiber bundles closely analogous to the one known for the
 corresponding dual jet bundles that appear in the multisymplectic and poly%
 symplectic approach to first order classical field theories.
\end{abstract}

\begin{flushright}
 \parbox{12em}{
  \begin{center}
   Universidade de S\~ao Paulo \\
   RT-MAP-0802 \\
   \mydate
  \end{center}
 }
\end{flushright}

\newpage

\setcounter{page}{1}

\section{Introduction}

The quest for a fully covariant hamiltonian formulation of classical field
theory has a long history. In particular, the search for a first order
formalism analogous to that for classical mechanics in terms of concepts
from symplectic geometry has stimulated the development of new geometric
tools usually referred to as ``multisymplectic'' or ``poly\-symplectic''
structures whose real significance is emerging only gradually. In fact,
for many years there has not even been a convincing general definition,
although a standard class of examples in terms of duals of jet bundles
has long been known and widely used.%
\footnote{This situation has been strikingly similar to that in classical
mechanics before it was realized that symplectic manifolds, rather than just
cotangent bundles, provide an adequate framework if one wants to accomodate
phenomena such as half-integral spin within classical mechanics.} \linebreak
This defect has recently been overcome~\cite{FG}, and it has been realized
that both multi\-symplectic and polysymplectic structures (which are not
the same thing) play an important role in the formalism; in particular, a
multisymplectic structure always induces a special kind of polysymplectic
structure by means of a construction called the ``symbol''. But the fact
that both types come together to form a pair has been anticipated almost 20
years ago~\cite{CCI}, when it became apparent that the covariant hamiltonian
formulation of first order classical field theories requires the simultaneous
use of two types of ``multiphase space'' that we shall refer to as ``ordinary
multiphase space'' and ``extended multiphase space'', respectively.

To be more precise, let us briefly recall the cornerstones of the construction
of the two types of multiphase space for first order lagrangian field theories;
for more details, the reader is referred to~\cite{CCI,GIM,FR}. The starting
point is the choice of a fiber bundle~$E$ over the space-time manifold~$M$
called the configuration bundle because its sections represent the basic
fields of the theory at hand. Next, one takes the first order jet bundle
$JE$ of~$E$ to accomodate first order derivatives of these fields: this
is an affine bundle over~$E$ and is also the domain of definition of the
lagrangian. Besides, one also considers the linearized first order jet
bundle $\vec{J} E$ of~$E$: this is a vector bundle over~$E$ defined as
the difference vector bundle of $JE$. Finally, as in mechanics, one uses
appropriate versions of the Legendre transformation induced by the given
lagrangian to pass to the (twisted) affine dual $J\upostar E$ of $JE$ and
to the (twisted) linear dual $\vec{J}\upoast E$ of $\vec{J} E$: the former
is the extended multiphase space and the latter is the ordinary multiphase
space of the theory. Note that the former is an affine line bundle over the
latter and that the hamiltonian obtained from the given lagrangian through
Legendre transformation is not a function but rather a section of this
affine line bundle, so both of these multiphase spaces are essential
ingredients for defining the concept of a hamiltonian system in field
theory! Moreover, it is well known that $J\upostar E$ carries a naturally
defined multisymplectic form $\,\omega$. However, what does not seem to
have been so widely noticed is the fact that $\vec{J}\upoast E$ carries
a naturally defined polysymplectic form $\,\hat{\omega}$~-- even though
this form already appears explicitly in Ref.~\cite{Gu}. As has been shown
more recently~\cite{FG}, it can be derived from the multisymplectic form
$\,\omega$ on $J\upostar E$ by taking its symbol, which turns out to be
degenerate precisely along the fibers of the aforementioned affine line
bundle, and then passing to the corresponding quotient of $J\upostar E$
by the kernel of $\,\omega$, which is precisely $\vec{J}\upoast E$.
Note that this polysymplectic form $\,\hat{\omega}$ on $\vec{J}\upoast E$
is canonical, whereas the form $\,\omega_{\mathscr{H}}^{}$ on~$\vec{J}\upoast E$
obtained as the pull-back of~$\,\omega$ by means of a hamiltonian section
$\; \mathscr{H}: \vec{J}\upoast E \rightarrow J\upostar E \;$ is not,
since it depends on the choice of hamiltonian.%
\footnote{It should be noted that the form $\,\omega_{\mathscr{H}}^{}$
is closed and non-degenerate but not multisymplectic in the sense of the
definition given in Ref.~\cite{FG}.}

In terms of adapted local coordinates $(x^\mu,q^i,p\>\!_i^\mu,p)$ for
$J\upostar E$ and $(x^\mu,q^i,p\>\!_i^\mu)$ for $\vec{J}\upoast E$,
induced by local coordinates $x^\mu$ for~$M$, local coordinates $q^i$ for
the typical fiber~$Q$ of~$E$ and a local trivialization of~$E$~\cite{FR},
we have
\begin{equation} \label{eq:MSF1}
 \begin{array}{c}
  \mbox{extended multiphase space $J\upostar E$} \\[1mm]
  \mbox{adapted local coordinates $(x^\mu,q^i,p\>\!_i^\mu,p)$} \\[1mm]
  \mbox{multisymplectic form} \quad
  \omega~=~dq^i \,\smwedge\, dp\>\!_i^\mu \,\smwedge\, d^{\,n} x_\mu^{} \, - \;
           dp \,\,\smwedge\, d^{\,n} x
 \end{array}
\end{equation}
and
\begin{equation} \label{eq:PSF1}
 \begin{array}{c}
  \mbox{ordinary multiphase space $\vec{J}\upoast E$} \\[1mm]
  \mbox{adapted local coordinates $(x^\mu,q^i,p\>\!_i^\mu)$} \\[1mm]
  \mbox{polysymplectic form} \quad
  \hat{\omega}~=~dq^i \,\smwedge\, dp\>\!_i^\mu \,\smotimes\, d^{\,n} x_\mu^{}
 \end{array}
\end{equation}
where $p\,$ is (except for a sign) a scalar energy variable and $d^{\,n} x$
is the (local) volume form induced by the $x_{}^\mu$ while $d^{\,n} x_\mu^{}$
is the (local) $(n\!-\!1)$-form obtained by contracting $d^{\,n} x$ with
$\, \partial_\mu^{} \equiv \partial/\partial x_{}^\mu$:
\[
 d^{\,n} x_\mu^{}~=~i_{\partial_\mu^{}}^{} \, d^{\,n} x~.
\]
The same picture prevails in the general case if we replace adapted local
coordinates by Darboux coordinates; see~\cite{FG}. \emph{Extended multiphase
space is multisymplectic, ordinary multiphase space is polysymplectic.}

A crucial role in the development of the hamiltonian formalism is played by
the notion of a hamiltonian vector field. According to the picture outlined
above, this comes in two variants: a multisymplectic one and a polysymplectic
one. We shall deal with the two versions separately, beginning with the
pertinent definitions.

\section{The multisymplectic case}

According to Ref.~\cite{FG}, a multisymplectic fiber bundle of rank~$N$ can
be defined as a fiber bundle $P$ over an $n$-dimensional base manifold~$M$
equipped with a closed, non-degenerate $(n+1)$-form $\,\omega$ on its total
space~$P$ which (a)~is $(n-1)$-horizontal, i.e., such that its contraction
with any three vertical vector fields vanishes, and (b)~admits a multi%
lagrangian distribution, i.e., an isotropic vector subbundle~$L$ of the
vertical bundle~$VP$ of~$P$ of codimension $N$ and dimension $Nn+1$.
(It then turns out that $P$ has dimension $(N+1)(n+1)$.) \linebreak
Assuming this distribution to be involutive, which is automatic as
soon as $\, n \geqslant 3 \,$ but has to be imposed as a separate
condition when $\, n=2 \,$, Darboux's theorem assures that there
exist local coordinates, called canonical local coordinates or
Darboux coordinates, in which $\,\omega$ assumes the form
\begin{equation} \label{eq:MSF2}
 \omega~=~dq^i \,\smwedge\, dp\>\!_i^\mu \,\smwedge\, d^{\,n} x_\mu^{} \, - \;
          dp \,\,\smwedge\, d^{\,n} x~.
\end{equation}
Locally, $\,\omega$ is exact, i.e.,
\begin{equation} \label{eq:MSF3}
 \omega \; = \; - \, d\theta~,
\end{equation}
where $d$ denotes the exterior derivative, with
\begin{equation} \label{eq:MSF4}
  \theta~=~p\>\!_i^\mu \; dq^i \,\smwedge\, d^{\,n} x_\mu^{} \, + \;
           p \; d^{\,n} x~.
\end{equation}
The standard example is that of the extended multiphase space $J\upostar E$
mentioned above, for which $\,\omega$ is also globally exact, i.e., the
so-called multicanonical form $\theta$ in equations~(\ref{eq:MSF3}) and~%
(\ref{eq:MSF4}) is globally defined, and $L$ is the vector subbundle of~$VP$
generated by the vector fields $\partial/\partial p\>\!_i^\mu$ and $\partial/%
\partial p$, that is, the vertical bundle for the projection of $J\upostar E$
onto~$E$ (with respect to which $J\upostar E$ is a vector bundle).

Given this situation, we say that a vector field $X$ on~$P$ is
\emph{locally hamiltonian\/} if $i_X^{} \omega$ is closed, or
equivalently, if
\begin{equation} \label{eq:LMHVF} 
 L_X^{} \omega~=~0~.
\end{equation}
It is called \emph{globally hamiltonian\/} if $i_X^{} \omega$ is exact,
that is, if there exists an $(n-1)$-form $f$ on~$P$ such that
\begin{equation} \label{eq:GMHVF} 
 i_X^{} \omega~=~df~.
\end{equation}
In this case, $f$ is said to be a \emph{hamiltonian form associated with} $X$.
Finally, when $\,\omega$ is exact and given by equation~(\ref{eq:MSF3}), $X$
is called \emph{exact hamiltonian\/} if
\begin{equation} \label{eq:EMHVF} 
 L_X^{} \theta~=~0~.
\end{equation}
The main theorem states that these vector fields can be classified in terms
of their components with respect to canonical local coordinates, which are
given by the expansion
\begin{equation} \label{eq:EMPSVF} 
 X~=~X_{}^\mu \, \frac{\partial}{\partial x_{\phantom{i}}^\mu} \, + \,
     X_{}^i \, \frac{\partial}{\partial q^i} \, + \,
     X_i^\mu \, \frac{\partial}{\partial p\>\!_i^\mu} \, + \,
     X_0^{} \, \frac{\partial}{\partial p}~,
\end{equation}
whereas, locally, the hamiltonian form corresponding to such a vector field,
which is determined up to an arbitrary closed form, can be assumed to have an
expansion of the form
\begin{equation} \label{eq:EMPSHF} 
 f~=~f_{}^\mu \; d^{\,n} x_\mu^{} \, + \, {\textstyle \frac{1}{2}} \,
     f_i^{\mu\nu} \; dq^i \,\smwedge\, d^{\,n} x_{\mu\nu}~,
\end{equation}
where
\[
 d^{\,n} x_{\mu\nu}^{}~=~i_{\partial_\nu}^{} i_{\partial_\mu}^{} \, d^{\,n} x~.
\]
An easy calculation gives
\begin{equation} \label{eq:CONTM1} 
 \begin{array}{rcl}
  i_X^{} \omega \!\!
  &=&\!\! X_{}^\nu \; dq^i \,\smwedge\, dp\>\!_i^\mu \,\smwedge\,
          d^{\,n} x_{\mu\nu} \, - \,
          X_i^\mu \; dq^i \,\smwedge\, d^{\,n} x_\mu \, + \,
          X_{}^i \; dp\>\!_i^\mu \,\smwedge\, d^{\,n} x_\mu \\[2mm]
  & &\!\! \mbox{} + \,
          X_{}^\mu \; dp \,\,\smwedge\, d^{\,n} x_\mu \, - \,
          X_0^{} \; d^{\,n} x~,
 \end{array}
\end{equation}
and in the exact case
\begin{equation} \label{eq:CONTM2} 
 i_X^{} \theta~
 =~(p\>\!_i^\mu X_{}^i \, + \, p \, X_{}^\mu) \; d^{\,n} x_\mu \, - \,
   p\>\!_i^\mu X_{}^\nu \; dq^i \,\smwedge\, d^{\,n} x_{\mu\nu}~.
\end{equation}
These formulas constitute the starting point for the proof of the following
\begin{thm} \label{thm:HAMVFM}~
 A vector field\/ $X$ on~$P$ is locally hamiltonian if and only if its
 components\/ $X_{}^\mu$, $X_{}^i$, $X_i^\mu$ and $X_0^{}$ with respect
 to canonical local coordinates, as defined by equation~(\ref{eq:EMPSVF}),
 satisfy the following conditions:
 \begin{enumerate}
  \item the coefficients\/ $X_{}^\mu$ and\/ $X_{}^i$ are independent of
        the multimomentum variables $p\>\!_k^\kappa$ and of the energy
        variable $p$, with the coefficients $X_{}^\mu$ depending only
        on the local coordinates $x^\kappa$ of the base manifold~$M$
        as soon as $\, N>1$,
  \item the remaining coefficients\/ $X_i^\mu$ and\/ $X_0^{}$ can be
        expressed in terms of the previous ones and of new coefficients
        $f_0^\mu$ which are also independent of the multimomentum
        variables $p\>\!_k^\kappa$ and of the energy variable $p$,
        according to
        \begin{equation} \label{eq:LHAMMVF01}
         X_i^\mu~
         = \; - \, p \; \frac{\partial X_{}^\mu}{\partial q^i} \, - \,
           p\>\!_j^\mu \; \frac{\partial X_{}^j}{\partial q^i} \, + \,
           p\>\!_i^\nu \; \frac{\partial X_{}^\mu}{\partial x_{}^\nu} \, - \,
           p\>\!_i^\mu \; \frac{\partial X_{}^\nu}{\partial x_{}^\nu} \, + \,
           \frac{\partial f_0^\mu}{\partial q^i}~,
        \end{equation}
        (the first term being absent as soon as $\, N>1$) and
        \begin{equation} \label{eq:LHAMMVF02}
         X_0^{}~
         = \; - \, p \; \frac{\partial X_{}^\mu}{\partial x_{}^\mu} \, - \,
           p\>\!_i^\mu \; \frac{\partial X_{}^i}{\partial x^\mu} \, + \,
           \frac{\partial f_0^\mu}{\partial x^\mu}~.
        \end{equation}
 \end{enumerate}
 The components of the corresponding hamiltonian form $f$ are given by
 \begin{equation}
  f_{}^\mu~=~p\>\!_i^\mu X_{}^i \, + \, p \, X_{}^\mu \, + \, f_0^\mu~,
 \end{equation}
 and
 \begin{equation}
  f_i^{\mu\nu}~=~p\>\!_i^\nu X_{}^\mu \, - \, p\>\!_i^\mu X_{}^\nu~.
 \end{equation} 
 In addition, when $\,\omega$ is exact and given by equation~(\ref{eq:MSF3}),
 $X$ is exact hamiltonian if and only if the coefficients $f_0^\mu$ vanish.
\end{thm}

\noindent
This theorem has been explicitly stated in Ref.~\cite{FHR} and an explicit
proof of a more general theorem (where vector fields are replaced by multi%
vector fields) can be found in Ref.~\cite{FPR}.

\section{The polysymplectic case}

According to Ref.~\cite{FG}, a polysymplectic fiber bundle of rank~$N$ can
be defined as a fiber bundle $P$ over an $n$-dimensional base manifold~$M$
equipped with a vertically closed, non-degenerate vertical $2$-form
$\,\hat{\omega}$ on its total space~$P$ which (a)~takes values in (the
pull-back $\pi^* \hat{T}$ to~$P$ of) some given $\hat{n}$-dimensional
coefficient vector bundle $\hat{T}$ over~$M$ and (b)~admits a poly%
lagrangian distribution, i.e., an isotropic vector subbundle~$L$ of
the vertical bundle~$VP$ of~$P$ of codimension~$N$ and dimension~%
$N \hat{n}$. (It then turns out that $P$ has dimension $N \hat{n}
+ N + \hat{n}$.) Assuming this distribution to be involutive, which
is automatic as soon as $\, \hat{n} \geqslant 3 \,$ but has to be
imposed as a separate condition when $\, \hat{n}=2 \,$, Darboux's
theorem assures that given any basis $\, \{ \, \hat{e}_a \, / \,
1 \leqslant a \leqslant \hat{n} \, \} \,$ of local sections of
$\hat{T}$, there exist local coordinates, called canonical local
coordinates or Darboux coordinates, in which $\,\hat{\omega}$
assumes the form
\begin{equation} \label{eq:PSF2}
 \hat{\omega}~=~dq^i \,\smwedge\, dp\>\!_i^a \,\smotimes\, \hat{e}_a^{}~.
\end{equation}
Locally, $\,\hat{\omega}$ is vertically exact, i.e.,
\begin{equation} \label{eq:PSF3}
 \hat{\omega} \; = \; - \, d_V \hat{\theta}~,
\end{equation}
where $d_V$ denotes the vertical exterior derivative, with
\begin{equation} \label{eq:PSF4}
 \hat{\theta}~=~p\>\!_i^a \; dq^i \,\smotimes\, \hat{e}_a^{}~.
\end{equation}
The standard example is that of the ordinary multiphase space
$\vec{J}\upoast E$ mentioned above, with $\, \hat{T} = \bwedge^{n-1}
T^* M \,$ and $\, e_a^{} = d^{\,n} x_\mu^{}$, for which $\,\hat{\omega}$
is also globally vertically exact, i.e., the so-called polycanonical
form $\hat{\theta}$ in equations~(\ref{eq:PSF3}) and~(\ref{eq:PSF4})
is globally defined, and $L$ is the vector subbundle of~$VP$ generated
by the vector fields $\partial/\partial p\>\!_i^\mu$, that is, the
vertical bundle for the projection of $\vec{J}\upoast E$ onto~$E$
(with respect to which $\vec{J}\upoast E$ is a vector bundle).

Given this situation, we say that a vertical vector field $X$ on~$P$ is
\emph{locally hamiltonian\/} if $i_X^{} \hat{\omega}$ is vertically closed,
or equivalently, if
\begin{equation} \label{eq:LPHVF} 
 L_X^{} \hat{\omega}~=~0~.
\end{equation}
It is called \emph{globally hamiltonian\/} if $i_X^{} \hat{\omega}$ is
vertically exact, that is, if there exists a section $f$ of the vector
bundle $\pi^* \hat{T}$ over~$P$ such that
\begin{equation} \label{eq:GPHVF} 
 i_X^{} \hat{\omega}~=~d_V f~.
\end{equation}
In this case, $f$ is said to be a \emph{hamiltonian section associated with}
$X$. Finally, when $\,\hat{\omega}$ is vertically exact and given by equation~%
(\ref{eq:PSF3}), $X$ is called \emph{exact hamiltonian\/} if
\begin{equation} \label{eq:EPHVF} 
 L_X^{} \hat{\theta}~=~0~.
\end{equation}
The main theorem states that these vector fields can be classified in terms
of their components with respect to canonical local coordinates, which are
given by the expansion
\begin{equation} \label{eq:OMPSVF} 
 X~=~X_{}^i \, \frac{\partial}{\partial q^i} \, + \,
     X_i^a \, \frac{\partial}{\partial p\>\!_i^a}
\end{equation}
whereas, locally, the hamiltonian section corresponding to such a vector field,
which is determined up to (the pull-back to~$P$ of) an arbitrary section of~%
$\hat{T}$, can be assumed to have an expansion of the form
\begin{equation} \label{eq:OMPSHF} 
 f~=~f_{}^a \; \hat{e}_a^{}~.
\end{equation}
An easy calculation gives
\begin{equation} \label{eq:CONTP1} 
 i_X^{} \hat{\omega}~= \; - \, X_i^a \; dq^i \,\smotimes\, \hat{e}_a^{} \, + \,
                       X_{}^i \; dp\>\!_i^a \,\smotimes\, \hat{e}_a^{}~,
\end{equation}
and in the exact case
\begin{equation} \label{eq:CONTP2} 
 i_X^{} \hat{\theta}~=~X_{}^i p\>\!_i^a \; \hat{e}_a^{}~.
\end{equation}
These formulas constitute the starting point for the proof of the following

\begin{thm} \label{thm:HAMVFP}~
 A vector field\/ $X$ on~$P$ is locally hamiltonian if and only if its
 components\/ $X_{}^i$ and $X_i^\mu$ with respect to canonical local
 coordinates, as defined by equation~(\ref{eq:OMPSVF}), satisfy the
 following conditions:
 \begin{enumerate}
  \item the coefficients\/ $X_{}^i$ are independent of the multimomentum
        variables $p\>\!_k^\kappa$,
  \item the remaining coefficients\/ $X_i^a$ can be expressed in term of
        the previous ones and of new coefficients $f_0^a$ which are also
        independent of the multimomentum variables $p\>\!_k^\kappa$,
        according to
        \begin{equation} \label{eq:LHAMMVF03}
          X_i^a~
          = \; - \, p\>\!_j^a \; \frac{\partial X_{}^j}{\partial q^i} \, + \,
            \frac{\partial f_0^a}{\partial q^i}~.
        \end{equation}
 \end{enumerate}
 The components of the corresponding hamiltonian section $f$ are given by
 \begin{equation}
  f_{}^a~=~p\>\!_i^a X_{}^i \, + \, f_0^a~,
 \end{equation}
 In addition, when $\,\hat{\omega}$ is vertically exact and given by
 equation~(\ref{eq:PSF3}), $X$ is exact hamiltonian if and only if the
 coefficients $f_0^a$ vanish.
\end{thm}

\noindent
The proof of this theorem is entirely analogous to that of the previous one,
except that it is much simpler. The essence of the argument can already be
found in Ref.~\cite{Ka1,Ka2}, but the proper global context of the result
is not adequately treated there.

\section{Outlook}

The analogous problem of determining hamiltonian vector fields with respect
to the form $\,\omega_{\mathscr{H}}^{}$ on ordinary multiphase space mentioned
in the introduction has been addressed and solved in Ref.~\cite{Sa}, but the
results are somewhat complicated and not very enlightening. We now believe
this to be related to the fact that, according to the structurally natural
definition given in Ref.~\cite{FG}, $\omega_{\mathscr{H}}^{}$ is \emph{not}
multisymplectic.

One problem that, for the time being, remains open is to give a global,
coordinate independent formulation of the results of Theorems~1 and~2.
This question is presently under investigation.

\end{document}